\documentstyle[cjourps,psfig]{cjour}
\begin{document}



\authorrunninghead{R. Capuzzo--Dolcetta et al.}
\titlerunninghead{On the use of MIMD--SIMD platform}






\title{On the use of a heterogeneous MIMD-SIMD platform to
simulate the dynamics of globular clusters with a central massive
object}

\author{R. Capuzzo Dolcetta$^1$, N. Pucello$^{2,3}$, V. Rosato $^{2,4}$, F. Saraceni$^2$}
\affil{$^1$\it Phys. Dept., Universit\'a di Roma "La Sapienza", P.le A. Moro 2, 00185 Roma (Italy)\\
$^2$\it ENEA, Casaccia Research Center, HPCN Project, P.O. Box 2400, 00100 Roma (Italy)\\
$^3$\it CASPUR,  Universit\'a di Roma "La Sapienza", P.le A. Moro 2, 00185 Roma (Italy)\\
$^4$\it Istituto Nazionale di Fisica della Materia (INFM), UdR
Roma I}

%









\abstract{The dynamics of a large stellar (globular) cluster
containing N=128,000 stars has been simulated by a direct
summation (O(N$^2$)) method by using a heterogeneous platform.
Preliminary simulations have been carried out on model systems
with and without the presence,
 in their center of mass, of a black hole whose
 mass has been varied from $0.02$ to $0.1$ times the total mass
of the cluster. These simulations followed the evolution of the
globular cluster in order to describe its dynamics over an
interval of time sufficiently large respect to the internal
crossing time.  Computations have demonstrated that the platform
heterogeneity, allowing a very efficient use of the computational
resources, can be considered a key feature for sustaining large
computational loads. Our results show that the massive object in
the center of the cluster alters the surrounding star
distribution very quickly; the following evolution is much slower
as it occurs via two-body collisional relaxation.}

\keywords{self--gravitating N--body simulations; heterogeneous
MIMD--SIMD platform }

\begin{article}



\section{Introduction}
Although the main trend of high performance computing seems to
focus on the assembly of larger and larger Massively Parallel
Platforms (MPP) (e.g. the US ASCI platforms \cite{ref:ASCI}), the
design, the realization and the deployment of specialized (or
dedicated) hardware is at the center of a renewed interest,
particularly in the field of scientific applications.

In the recent years, several scientific domains have received a
substantial impulsion by the use of specialized (or dedicated)
hardware: the Lattice Quantum Chromodynamics (LQCD) model, thanks
to the platforms realized in the frame of the APE
\cite{ref:Bertoloni1, ref:Bertoloni2} and the
 Columbia projects \cite{ref:Columbia}, has significantly increased its prediction
power. Computational astrophysics has received significant
benefits by the use of
 the dedicated platform GRAPE (GRAvitational Pipe) \cite{ref:Makino} which is
able to sustain a computational power of the order of
Teraflop/sec. in the gravitational O(N$^2$)
 calculations. More recently, the use of more complex N-body codes requiring
a substantial amount of O(NlogN) calculations has triggered the
development of new dedicated hardware, built by exploiting new
programmable devices (FPGA), used in combination with the GRAPE
machine (AHA-GRAPE project \cite{ref:AHA}).

Other dedicated and specialized hardware devices have been also
conceived to improve the
 sustained computational power of numerical modeling in the field of 3-dimensional
 Ising models \cite{ref:Bode}, \cite{ref:Tal}, in that of molecular orbital calculation
 in quantum chemistry \cite{ref:MOE} and in molecular dynamics simulation of complex systems
(e.g. biological matter) \cite{ref:DEMOS}. This list, far from
being exhaustive, puts together both dedicated (non-programmable)
and specialized (programmable) devices. Whereas dedicated
architectures cannot be deployed in application fields different
from those for that they have been conceived, specialized
hardware (like e.g. the
 APE platform) can be usefully deployed also in different application fields.

In order to take the most from these platforms, their power is
often exploited in conjunction with a host platform (a simple
workstation or a general
 purpose parallel platform). In this configuration, the
 specialized/dedicated platform can be regarded as a
smart co-processor of the host to which specific parts of the
computation can be allotted. The resulting hardware is, thus, the
{\it heterogeneous} sum of general--purpose and
specialized/dedicated machines; it allows to produce very
 efficient calculations as
different (in terms of algorithmic structure) parts of the code
can be
 allotted to the parts of the architecture most
 suited to map their specific computational complexity.

This work has addressed two different issues.
\par\noindent The first is concerned with the astrophysical problem of simulating,
with a reliable model, the dynamics of a large stellar cluster
whose central region hosts a massive object (like a black hole).
The aim is to apply these numerical models, which are approximated
just in what regards the numerical time integration scheme, to the
evolution of the cluster mass distribution over an interval of
time of the order of ten crossing times (the {\it crossing} {\it
time} $t_c$= 2R/$<v>$, is the time required by the typical object
of the system to cross the system, i.e. to cover a length scale
equal to twice the initial radius of the system $R$, with a
velocity $<v>$ equal to the average star's velocity). Actually,
 we estimate this interval of time as sufficient to sum up collective
effects even if it is still short respect to the two-body
relaxation time that is the \lq graininess \rq --dependent time
scale.

The representation of a large self-gravitating system with a
particle model is, indeed, limited by the number N of particles
that can be numerically handled; astrophysical
 self-gravitating systems span a range of N from $2$ to $10^{12}$ (see Table 1). The direct
 evaluation (i.e. that obtained by the sum of the contribution to the force on an  object
 of the system by all the others without approximations with the exception of a possible smoothing of the
 interaction potential) of the N--body forces implies a restrictive limit to the number of bodies
 which can be treated (being the algorithm of a O(N$^2$) scaling).
  At present, the largest value of N that can be approached with such methods,
 on large computational
   platforms, is N$\simeq$ $10^{5}$. This means that a one-to-one representation of stars with
    simulating particles is now possible only for stellar systems up to a typical (not too populous)
     globular cluster (see Table 1). Larger systems can be
     treated with approximate methods such as {\it tree codes}
     \cite{ref:BarHut}, {\it particle--mesh}(PM) and
{\it particle--particle, particle--mesh}(P$^3$M) \cite{ref:P3M}
algorithms.

\begin{table}[htb]
\caption{The table provides, for several astrophysical systems,
the corresponding range of the number of
  stars in the system, the numerical approximation commonly used by  simulations and the two--body relaxation time
  of the object in units of crossing time. The value of the relaxation time $t_{rel}$ has been expressed in
terms of the
  crossing time $t_c$ by means of the relation $t_{rel}\approx 0.1 \frac{N}{ln N} t_c $.}
\begin{tabular*}{\textwidth}{@{\extracolsep{\fill}}cccc}
\hline
  Stellar system     &   N                  & Numerical approximation    & $t_{rel}/t_c$  \\
  \hline
  OB associations    &  10-100              &  direct                    &  $1-5$                 \\
  Open clusters      &  100-1000            &  direct                    &  $5-30$                 \\
  Globular clusters  &  10$^5$-10$^6$       &  direct                    &  $10^3-10^4$\\
  Dwarf galaxies     &  10$^8$              & tree-algorithm,PM,P3M      &  $10^7$          \\
  Galaxies           &  10$^{10}$-10$^{12}$ & tree-algorithm,PM,P3M  &     $10^8-10^9$             \\
  \hline
\end{tabular*}
\end{table}

The second aim of this work is to focus on the key role which
could be played by heterogeneous platforms to efficiently cope
with complex computational problems. In this case, an
heterogeneous platform, made up by connecting a specialized
architecture to a general--purpose platform, is used in a
computationally intensive computational problem. The efficient
use of the platform (induced by a suitable mapping of the
algorithm onto the machine architecture and the use of a smart
communication scheme)
 has allowed to reach a sustained computational
power of the order of 10 Gflop/sec, necessary to perform
extensive calculations.

The scheme of the paper is the following. The next section
 provides a brief outlook of the heterogeneous platform used to
perform computations. The definition of the physical model
considered for the computations is the object of the third
section, which also contains a schematic layout of the techniques
used to implement the calculations on the computing architecture.
The fourth and the fifth  sections  are devoted to the
presentation and the discussion of the results. In these sections
 both the scientific relevance of the resulting data and the fate of the
 heterogeneous computing in complex scientific calculations are discussed.

\section{Layout of the computational platform}
The heterogeneous platform used for computations is a MIMD-SIMD
\cite{ref:nota} platform,
 called PQE1 (Fig.1), realized in the frame of an industrial program by ENEA and
QSW (an italian company of the Finmeccanica group)
\cite{ref:PQE1}. The PQE1 is a platform where the flexibility and
the operability of a MIMD platform with a distributed--memory
architecture are coupled to the power
  and the efficiency of SIMD machines.
We assume that most algorithms arising in scientific applications
are mainly expressible through synchronous programs with
synchronous
 communications. These execute the same instruction on a set of different data which can be easily
  mapped onto a data parallel structure with regular patterns of memory access. Under these assumptions,
 it is reasonable to allot those parts to the SIMD machine,
   leaving the remaining tasks of the computation to be executed on the MIMD part.
The PQE1 platform consists of a MIMD general--purpose platform,
acting as a docking unit and of 7 SIMD machines.

The MIMD system is a a Meiko/QSW CS-2 platform with 8 nodes, each
of them based on a dual--HyperSparc processor at 125 MHz,
connected in the SMP configuration by a Meiko/QSW proprietary
network based on the Elan/Elite devices
   and implementing a multistage interconnection network with Fat Tree topology and point-to-point
    bandwidth of 100 Mbyte/sec. This platform offers
 a peak speed of 2.18 Gflops/sec. and has abuot 2 Gbytes of addressable RAM.

 The SIMD platforms come from the APE project (APE100 \cite{ref:Bertoloni1, ref:Bertoloni2})
 and have been produced by QSW with the commercial name of Quadrics.
 Theinstalled platforms
  have the following characteristics: two of them (called QH4) have 512
Floating Point Units (FPU) arranged as an
 (8x8x8) 3D torus and 5  of them (called QH1) have 128 FPU's , arranged as an (8x4x4) 3D torus. Each
 FPU is based on a custom VLIW processor, has clock frequency $\nu$=25
MHz and is able to terminate
   a 'normal operation' A*B+C ( where A, B and C are IEEE 754 standard, single precision, real numbers)
   every clock cycle. Each processor thus executes two floating point
    operations in one clock cycle (when the pipeline is full) and has a peak speed of 50 Mflops/sec.
   Each FPU
      is connected to a data memory of 4Mbytes and has an internal register file (RF) with 128
      registers.
      Each clock cycle the processor is able to read two operands from RF and write one result to RF.
      Communications with other adjacent FPU's, connected in the north, south, east, west, up and down
       directions are synchronous and memory mapped. The interprocessor communication bandwidth is 12.5 Mbyte/sec,
        so the 512 (128) processor configuration has an aggregate bandwidth of 6.4 (1.6) Gbyte/sec and a
        peak speed of 25.6 (6.4) Gflops/sec.

The SIMD APE100/Quadrics machines are connected to the MIMD
system through an HiPPI (High Performance Parallel Interface)
  channel, which provides a bandwidth of 20 Mbyte/sec. Each MIMD node is connected to a different SIMD
  platform, as in the scheme of the complete prototype reported in Fig.1.
Looking at previous data, it is clear that the machine is
strongly unbalanced, having the
 most of computational and communication speed in the SIMD part.

The rationale for the design of such a prototype platform is
related to the necessity of having a complex computing platform
whose complexity mirrors that of the computational codes used in
scientific applications. These latter are, in fact, formed by a
sequence of tasks, each of them characterized by a {\it task
granularity} $g_t$ given by

\begin{equation}
g_t = \frac{N_{op}}{D_{I/O}}
\end{equation}

where $N_{op}$ is the number of floating point operations to be
performed and $D_{I/O}$ is the I/O data to be processed in the
task. Analogously, it is possible to define a {\it machine
granularity} $g_m$ given by

\begin{equation}
g_m= \frac{F_{op}}{B}
\end{equation}

where $F_{op}$ is the computational power of the platform and B
its I/O bandwidth. It has been proven \cite{ref:cancun} that an
efficient implementation of the task on a given computational
platform implies that

\begin{equation}
g_m < \eta g_t
\end{equation}

where $\eta$ is a suitable factor $0<\eta<1$ expressing the
efficiency of the task implementation. For this reason, an
efficient implementation of a complex computational code implies
the simultaneous presence of a complex computational structure
able to satisfy the requirement of eq.(3) for each task composing
the code. The complexity of the PQE1 platform is thus exploited by
allotting the high-granularity tasks to the SIMD part and the
low--granularity and the pre/post processing tasks to the MIMD
machine.

\section{Model and computations}
We represent the stellar system as a set of N=128,000
point-masses (stars)
 interacting via the classical pair gravitational potential

\begin{equation}
V(r_{ij}) =- G\frac{m_im_j}{r_{ij}}
\end{equation}

where $r_{ij}$ is the distance among the i-th and j-th star of
the system. The stellar masses are assumed all equal. To avoid
force divergence, the interaction potential $V$ has been smoothed
by substituting
 $r_{ij}$ with $r'_{ij}$  =  $r_{ij}+\epsilon$.
The introduction of the smoothing parameter $\epsilon$ affects
the dynamics on that length scales. This distance, in general, is
set to smaller values than the average interparticle distance, in
order to ensure an acceptable level of approximation of the
system's dynamics.

When a massive object is present in the system (as a massive black
hole) another length scale is naturally introduced in the
problem: the tidal radius
$r_t$=$(\frac{m_{BH}}{2m})^{\frac{1}{3}}R_*$. A star of mass $m$
and radius $R_*$ approaching a black hole (of mass $m_{BH}$)
  at a distance smaller than r$_t$ is destroyed by the strong tidal deformation and, consequently,
   its mass goes to increase that of the
 black hole with a positive feedback on the tidal radius.
  Our computational code takes into account the possibility that a star
 is swallowed by the massive object which accordingly increases its mass and its r$_t$.
 However, r$_t$ is usually much smaller than the typical interstellar distance so that $m_{BH}$
 remains almost unchanged throughout the simulations.

The initial spatial distribution of the N-stars of equal masses
has been sampled by a spherical Plummer distribution (see \cite
{ref:Plummer})

\begin{equation}
\rho (r) = \frac{\rho_0}{[1+(\frac{r}{r_c})^2]^{\frac{5}{2}}}
\end{equation}

where the central density $\rho_0$ and the "core" radius $r_c$
are free parameters. The initial stars velocities have been
obtained self-consistently, from the velocity distribution
function that generates the Plummer's density law, i.e.

\begin{equation}
 f(\bf r,\bf v) = \begin{cases}  
{(-E)^{7/2}, E\leq 0 \cr 
0, E>0 }
 \end{cases}
\end{equation}

where $E=1/2v^2 - \phi(r)$ is the individual star's energy in the
(spherical) potential $\phi (r)$ given by the Plummer mass
distribution. The star velocities distributed according to eq. (6)
are scaled to have a virial ratio $Q=2T/|\Omega|=0.98$ ( where T
and $\Omega$ are the kinetic and the gravitational energies of
the system) where the equilibrium value is 1; the inclusion of a
black hole with mass $0.02$ and $0.1$ of the total star mass at
the center of mass changes the virial ratio to $Q=0.87$ and
$Q=0.75$, respectively, inducing a stronger gravitational
collapse.

The equations of motion have been integrated  by using the
central difference Verlet--scheme \cite{ref:Verlet} with a fixed
time--step . This scheme allows an accuracy of order $\Delta t^4$
in the positions and $\Delta t^2$ for the velocities.
   The time step $\Delta t$ has been empirically
selected by requiring the relative error of the total energy to
be smaller than 10$^{-4}$ per time step. It results that $\Delta
t$ is a fraction of the system's crossing time, $t_c$. After
having fixed the value of $\epsilon$, several simulations have
been carried out, with different values of $\Delta t$, up to the
value which allowed to obtain the relative error of the total
energy in the desired range.

The part of the PQE1 platform selected for the calculations
consists of a single MIMD node and  a QH4 platform. The MIMD node
has been used to perform
 data initialization, the evaluation of the component of the star forces arising from the interaction
with the black hole, the solution of the equations of motion and
the evaluation
 of the relevant physical quantities (total and potential energy, virial ratio, Lagrangian radii,
 mass and velocity distributions). The O(N$^2$) force-loop has been, in turn,
  implemented on the QH4 platform. For this purpose, a number of stars N$_p$=N/n$_p$
  (where n$_p$=512 is
the number of FPU of the QH4 platform) has been allotted to each
FPU. The force calculation has been performed by making use of a
recently developed hyper-systolic
 algorithm \cite{ref:Lippert} in order to exchange data among the processors. This
scheme allows reduction of the redundancy  typical of the
systolic algorithm \cite {ref:Petkov},
 thus allowing to save a substantial amount of computing time.

In the systolic-type algorithms, there is a regular sequence of
calculation and communication steps. In our case, after the
allocation of equal groups of $N_p$ stars on each FPU, the code
evaluates the N--body forces within each group. A copy of the
positions of each group is then moved, from the initial to the
next neighbor FPU and, there, interacted with the resident group
of stars. The process is repeated until each group of stars has
visited each other FPU.
 On each FPU is thus cumulated the force acting on the resident group of stars.

The O(N$^2$) problem requires, given $N$ data elements being
distributed among $p$ processors, a number $O(Np)$ of
inter-processor communications, i.e. $O(N)$ communications per
processor.  As this often leads to severe performance
bottlenecks, it is mandatory to search for new algorithms that
are able to reduce inter-processor communications to a number
$O(N)$ per processor.

The hyper-systolic algorithm, recently introduced, has the
potential to decrease inter-processor communications for
computational problems of the type mentioned above to a number
$O(N^{\frac{1}{2}})$ per processor.  This is achieved by a
1-dimensional systolic mapping of the data onto the processing
elements and the use of shortcut inter-processor communication
(hyper-connections) together with a replicated systolic mapping
of data.

The basic feature of hyper-systolic computing is a sequence of
shifts called HS-base which represent the hyper-systolic strides
along the 1-dimensional chain to be performed in the
hyper-systolic parallel calculation.  The direct computation of
{\em optimal} bases is exponentially expensive, with the number of
processors $p<64$. For larger numbers of processors, hitherto,
aside to the so-called {\em regular} bases, new HS-bases have
been determined \cite{ref:PalazzariLi} with a search technique
based on Simulated Annealing.

The implementation of the hyper-systolic communication scheme on
the N$^2$ calculation on the 512-node Quadrics QH4, has allowed
to achieve about 20 \% improvement in inter-processor
communication compared to the regular base implementation.

The chosen implementation strategy allows to exploit a double
level of parallelism:
 the first, consisting of the parallel implementation of the interstellar
 force calculation
in the SIMD platform by the hyper-systolic algorithm, the second
one related to the concurrent black hole-stars force calculation
performed by the host part during the time spent by the SIMD
machine to produce the interstellar force.

The machine performance was such to allow an execution time
$t_{ex}$=120 sec. for the calculation of the single time--step of
the cluster dynamics, by using the mentioned PQE1 partition.
About 20$\%$ of this time is related to different communication
actions (to/from the SIMD machine and within the SIMD machine, in
the hyper--systolic loop), the residual time being spent for the
number--crunching activity. The efficiency of the SIMD code (in
terms of efficiency of use of the single FPU resources) was
around 30$\%$, thus leading to a sustained computational power,
in the SIMD part, of about 9 GFlops/sec.

\section{ Computational details and results}
Three model systems have been simulated, all of them containing
N=128000 masses. The first one represents the globular
 cluster without any massive object; this system has been retained as reference
 and its behavior compared to that of the other systems. The second system
contains a black hole of mass $m_{BH}$=0.02 (in units of the total
star mass) located at the center of mass of the cluster. The
third system contains a black hole
 with $m_{BH}$=0.1.
We express the model parameters taking as units of length and
mass the initial radius of the system and its total mass,
respectively.
  Moreover,
setting G=1 implies that t$_c$ is the time unit. In these units,
the value of $\epsilon=R/500=0.002$ that we choose is much smaller
than the average inter--particle distance (which is $\propto
n^{-{1/3}}$, with $n$ average number density); this guarantees an
acceptable spatial resolution and keeps the newtonian behavior of
the interparticle force. In the case of the unperturbed cluster
(without the black--hole) a $\Delta t$=$t_c/250$ has been used.
The constraint imposed to the numerical scheme to ensure an
energy conservation of at
 least 10$^{-4}$ per time step implied the use of
 a smaller time step ($\Delta t$=$t_c/500$) when a black hole
 is present.
In all cases, the systems have been simulated during a time
interval of 10 crossing times.

The first reported quantity is the virial ratio $2T/|\Omega|$ of
the system (Figs. 2, 3 and 4). This quantity provides a first
picture of the global dynamics in the cluster driven by the
initial unbalance between kinetic and potential energy. The
virial ratio is strongly perturbed by the presence of the massive
object. This produces a strong driving force which brings the mass
distribution to a virial equilibrium that is slightly different
from $2T/|\Omega|=1$ due to the non--homogeneity of the
interaction potential function induced by the presence of the
softening parameter $\epsilon$. For the most massive black hole
here considered ($m_{BH}=0.1$) the virialization is achieved more
rapidly.

The virialization process is due to a violent relaxation that
occurs as a consequence of a series of damped (non--adiabatic)
oscillations of the system (see Fig. 5).

The evolution of both the spatial and projected density radial
distributions are shown in Figs. 6 to 9 where the initial and
final ($t=10 t_c$) profiles are reported. The time evolution
leads to a steepening of the profile toward the cluster center ($r
< 0.1$) more pronounced at larger black hole mass and to the
population of a \lq halo\rq ~($r > 1$) of high velocity stars.
Both the external profile and the inner one (the latter evolving,
of course, just in presence of the central
 compact object) are attained in a short time, of the order of one crossing time.

The slope of the space density inner distribution is for
$m_{BH}=0.1$ at $t=10 t_c$, $-0.92$, i.e. intermediate between
the values $-3/2$ and $-1/2$ characteristic, respectively, of the
density cusps analytically (and approximately)
 evaluated for the distribution of stars bound to a black hole and of those
 unbound
 \cite{ref:BinTre}.
We note that the slope of the halo region closely matches the
expected $r^{-2}$ profile (see, for instance, \cite{ref:ShaLig}).

The dynamical evolution of the model clusters is also
characterized by the appearance of some anisotropy in the
velocity distribution in the external regions. This has been
characterized by the evaluation of the anisotropy parameters

$$\beta_\theta = 1 - {<v_\theta^2>\over <v_r^2>}$$
$$\beta_\phi = 1 - {<v_\phi^2>\over <v_r^2>}$$

(the $<.>$ averages  represent the dispersions of the polar
components of the velocity) as functions of the radial distance
(see Figs. 10, 11 and 12). Being
 the initial distribution function given by Eq. (6) only dependent on the energy, the
initial velocity distribution is obviously isotropic, and both
$\beta_\theta$ and $\beta_\phi$ are zero (the three polar
components of the velocity being the same, in the average). The behaviour of
$\beta_\theta$ and $\beta_\phi$ is very similar for we expect the evolution
keeps the velocity vector invariant for rotation
 around the plane orthogonal to the radial direction. We have thus plotted in
 Figs. 10 to 12 just $\beta_\theta$. It is evident, in all the cases studied, a rapid
increase of $\beta$ in the outer regions, where, indeed, orbits
are more radially pointed due to the expansion of the stellar
halo. Moreover, we note that the more massive is the central
object the more evident is the negative minimum of the anisotropy
parameter around the initial boundary. In the case $m_{BH}=0.1$,
the transverse components of the velocity dominate over the radial
ones so to have $\beta \simeq -1$; this means that a quick effect of the
instability caused by the massive object is that the transverse
''temperature'' doubles the radial. This seems a transient effect
because, as the evolution goes on, the radial dependence of the
$\beta$ parameter tends to be rather insensitive to the  mass of
the central object. At $t=10t_c$, the isotropy is partially
restored in the average, being still evident a radially biased
distribution of orbits in the outskirts.

Finally, Fig.13 is a snapshot, at $t=10t_c$, of the star
distribution projected onto a coordinate plane of the model
cluster having the $m_{BH}=0.1$ black hole at its center.

\section{Discussion and conclusions}
Concerning with the main questions which have stimulated this
work, it is possible to draw the following conclusions:
\begin{itemize}
\item{the {\it heterogeneity} of the PQE1 platform
 can provide significant
benefits also in the gravitational N-body calculations, although
the structure of the computational problem is very different from
that for which the SIMD platform Quadrics/APE100 has been
conceived. This result confirms the relevance
 of the relation of eq.(3) which should be considered as the most important
constraint to be fulfilled in the mapping of a code onto a
computational platform.

The reported data confirms, in fact, that the increase of
computational efficiency provided by heterogeneous platforms
(where different architectures can coexist and provide an ideal
computational frame to execute different parts of the complex
computing
 codes characterized by different algorithmic complexity) can be seen as a
key feature for high performance computing whose benefits should
be more attentively searched and exploited.

The achieved sustained computational power allows to perform the
simulation of the dynamics of one crossing time (with N=128,000)
in a wall clock time of the order of $6 \cdot 10^4$ sec.; this
figure allows to perform simulations of the order of tens of
crossing times. The new SIMD machine, derived from the INFN
project APEmille and constituting the follow--up of the
Quadrics/APE100 platform, will enhance the computational power of
a factor ten. This will permit to study the evolution of the
cluster over a time of the order of hundreds of crossing times,
i.e. of the order the two--body relaxation time.}

\item{from the astrophysical point of view, this work has an
introductory relevance to the \lq direct \rq ~ study of the dynamics
of populous stellar systems (up to the scale of globular
clusters) whose internal density vries over a range of scales such that approximations in the interaction potential (as multipolar expansios, solution of Poisson's equation on a
grid, etc.) are of questionable reliability. By the way, this work
has shown how the inclusion of a massive object in a
quasi--equilibrium model of a globular cluster, represented as a
set of N=128,000 stars interacting with the \lq exact \rq~
potential (even if smoothed on a scale smaller than the average
interstellar distance), has consequences on the overall dynamics.
It induces a violent relaxation within few crossing times,
corresponding to an evolution of the star radial distribution
toward a central peaked distribution close to $\rho \propto
r^{-1}$ in the case of the most massive black hole considered
$m_{BH}$=0.1. Another effect caused by the presence of the
massive object is the loss of isotropy in the velocity
distribution, both in the outer regions ,where radial orbits dominate, and in an intermediate zone where, on the other side, transverse component of the velocity prevail.  Thenner region stands isotropic throughout the evolution.}
\end{itemize}

We have stored the phase--space trajectories of the stars all
along the simulated time. These data will allow a deeper
investigation of the dynamical processes occurring at the early
stages of the evolution and a careful characterization of the
average properties of the stellar phase--space distribution.







\begin{acknowledgments}
The authors kindly acknowledge discussions and suggestions from
 R. Spurzem and M.Hemsemdorf (ARI, Heidelberg), P. Miocchi (Univ. of Roma) and P. Spinnato (DAS, Amsterdam).
 P. Palazzari and M. Coletta (ENEA, Roma) are gratefully acknowledged for their help in the
 implementation of the hyper-systolic algorithm. Part of this work
has been supported by a INFN/PQE2000 grant to one of us (NP),
performed under the frame of the PQE1 project (ENEA--QSW) and
partially supported by CASPUR.
\end{acknowledgments}


\newpage
\authorrunninghead{R. Capuzzo--Dolcetta et al.}
\titlerunninghead{On the use of MIMD--SIMD platform}
\begin{figure}[ht]
       \noindent
       \psfig{file=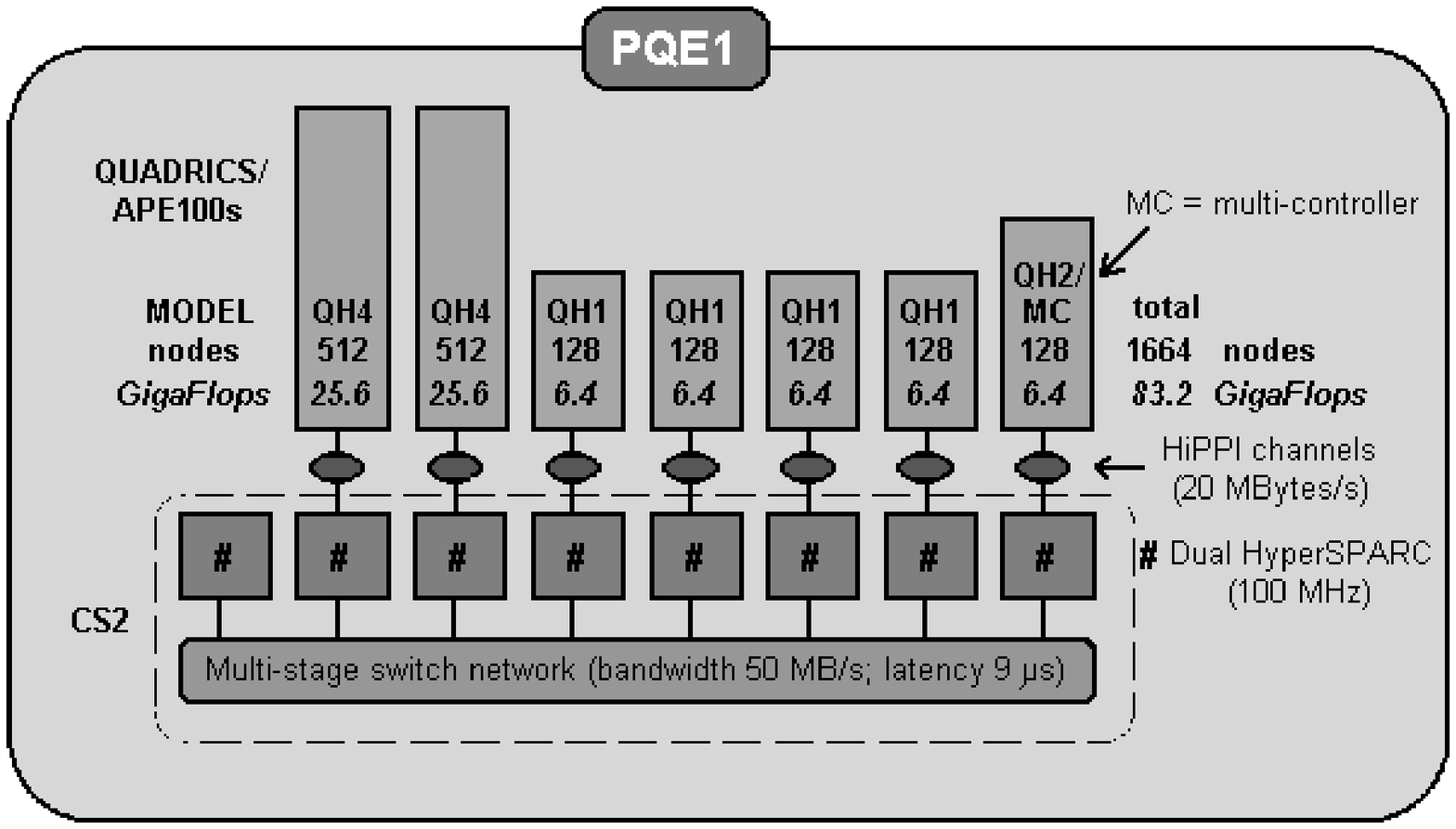,width=12.0cm}
       \vspace{0.5cm}
       \label{fig1}
\caption{Schematic layout of the PQE1 platform. The proprietary
multi-stage switch network of the CS2 platform has a fat tree
topology.
 Aside to the Hippi channel, there is a Transputer-based link which allows
communications to occur between the CS2 node and the connected
SIMD platform.}
\end{figure}
\newpage
\authorrunninghead{R. Capuzzo--Dolcetta et al.}
\titlerunninghead{On the use of MIMD--SIMD platform}
\begin{figure}[ht]
       \vspace{0.0cm}
       \noindent
\psfig{file=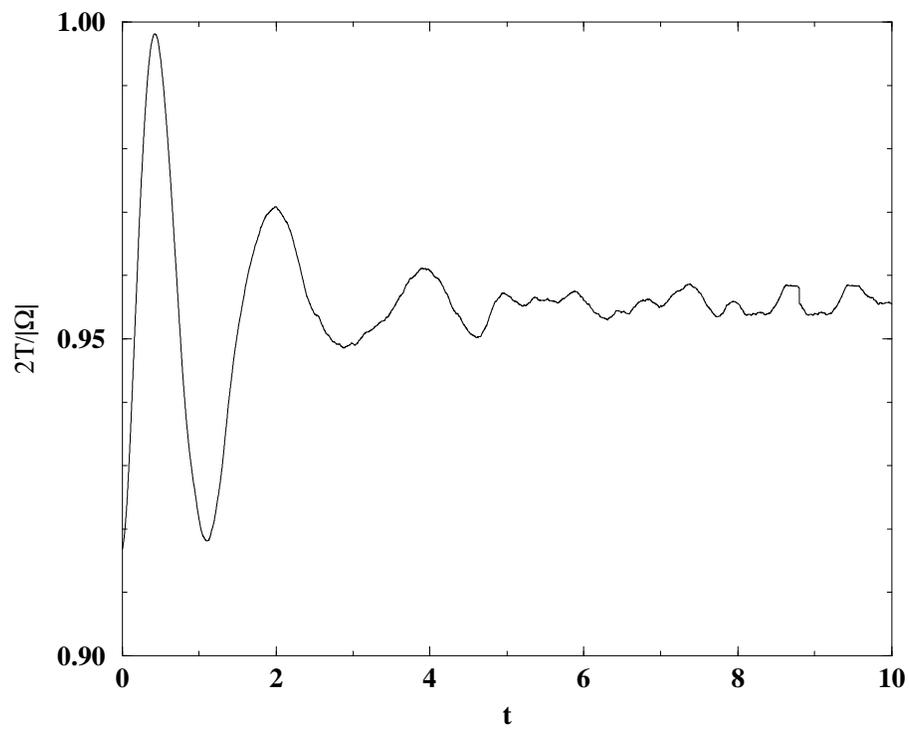,width=12.0cm}
       \vspace{0.5cm}
\label{fig2} \caption{Virial ratio, $\frac{2T}{|\Omega|}$, as a
function of time, for the system without the black--hole. Time is
expressed in units of crossing time $t_c$.}
\end{figure}

\newpage
\authorrunninghead{R. Capuzzo--Dolcetta et al.}
\titlerunninghead{On the use of MIMD--SIMD platform}
\begin{figure}[ht]
       \vspace{0.0cm}
       \noindent
\psfig{file=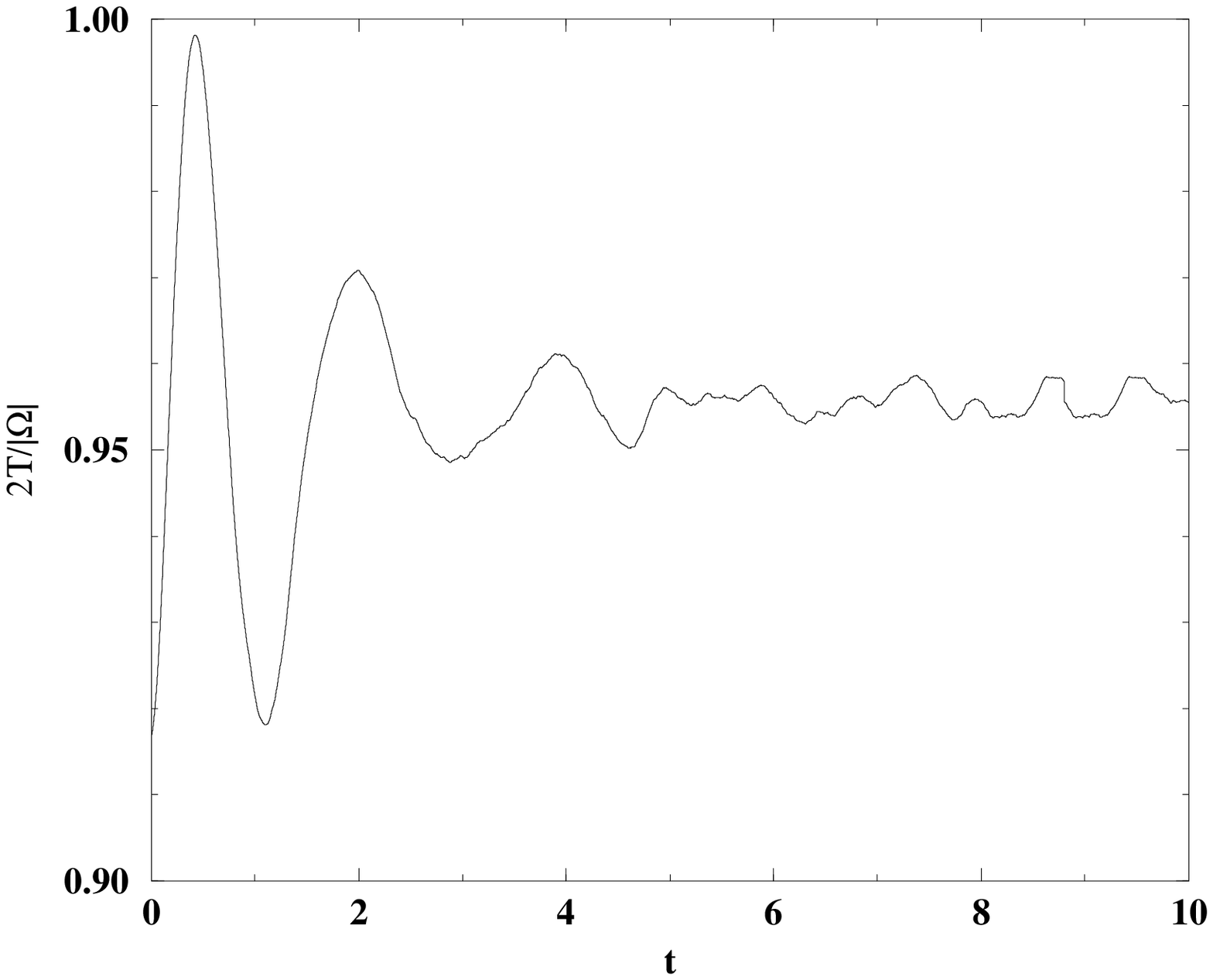,width=12.0cm}
       \vspace{0.5cm}
\label{fig3} \caption{Virial ratio, $\frac{2T}{|\Omega|}$, as a
function of time, for the system with the black--hole of mass
$m_{BH}$=0.02. Time is expressed in units of crossing time $t_c$.}
\end{figure}

\newpage
\authorrunninghead{R. Capuzzo--Dolcetta et al.}
\titlerunninghead{On the use of MIMD--SIMD platform}
\begin{figure}[ht]
        \vspace{0.0cm}
        \noindent
        \psfig{file=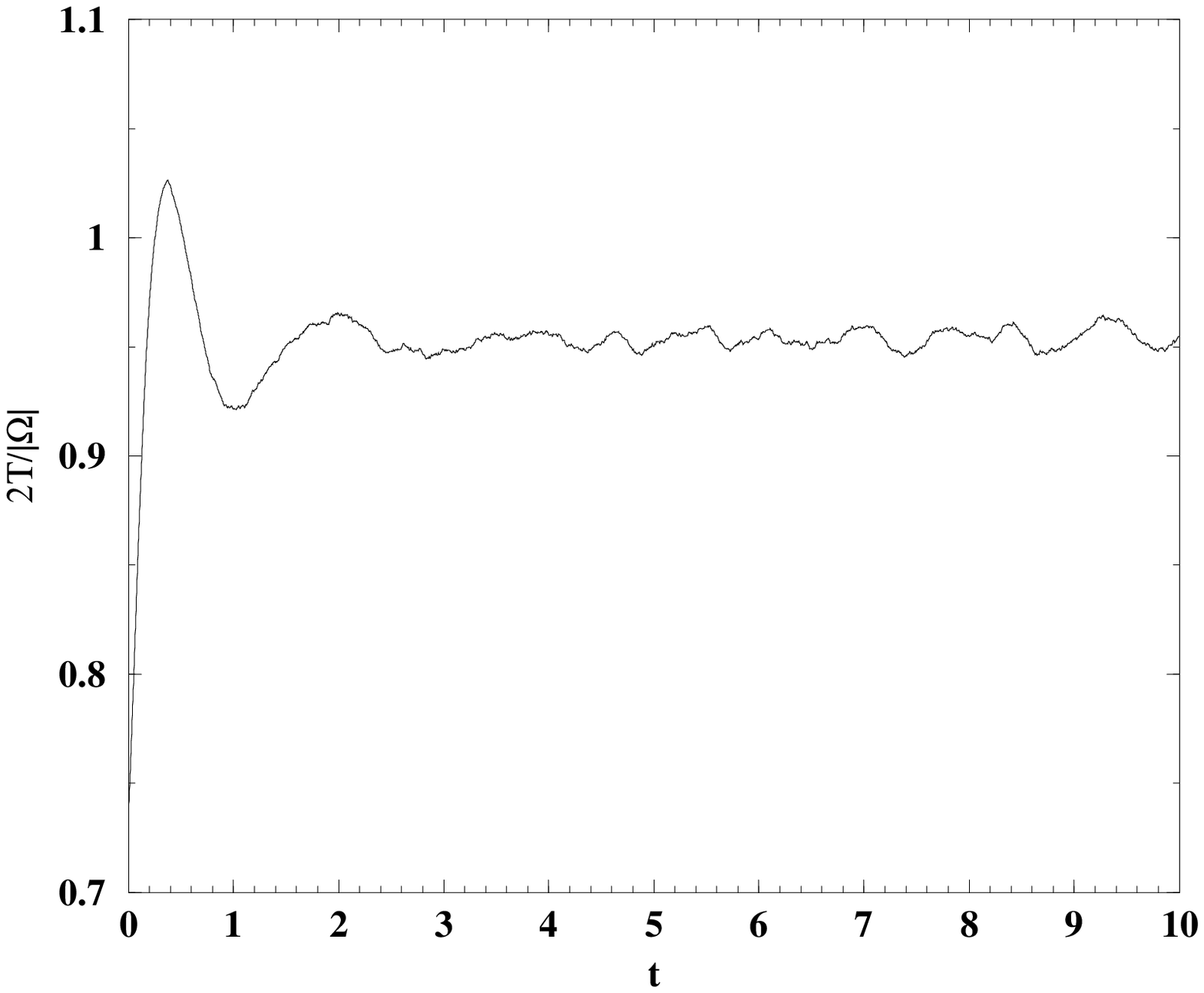,width=12.0cm}
        \vspace{0.5cm}
\label{fig4}
 \caption{Virial ratio, $\frac{2T}{|\Omega|}$, as a
function of time, for the system with the black--hole of mass
$m_{BH}$=0.1. Time is expressed in units of crossing time $t_c$.}
\end{figure}

\newpage
\authorrunninghead{R. Capuzzo--Dolcetta et al.}
\titlerunninghead{On the use of MIMD--SIMD platform}
\begin{figure}[ht]
        \vspace{0.0cm}
        \noindent
        \psfig{file=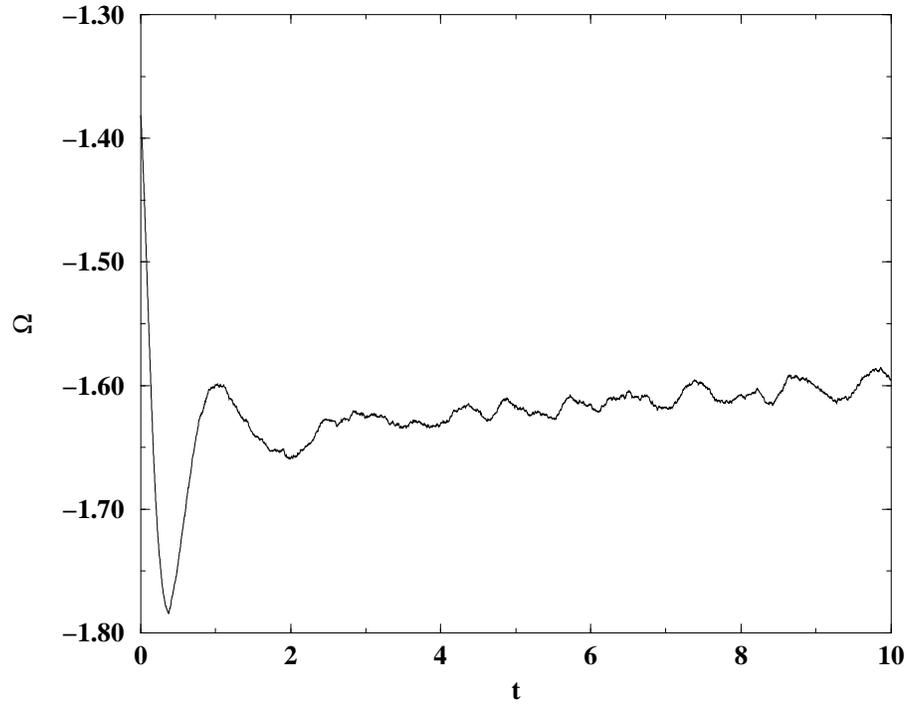,width=12.0cm}
        \vspace{0.5cm}
\label{fig5}
 \caption{Total potential energy $\Omega$ vs. time for the system
 with the black--hole of mass
$m_{BH}$=0.1. Time is expressed in units of crossing time $t_c$.}
\end{figure}

\newpage
\authorrunninghead{R. Capuzzo--Dolcetta et al.}
\titlerunninghead{On the use of MIMD--SIMD platform}
\begin{figure}[ht]
       \vspace{0.0cm}
       \noindent
       \psfig{file=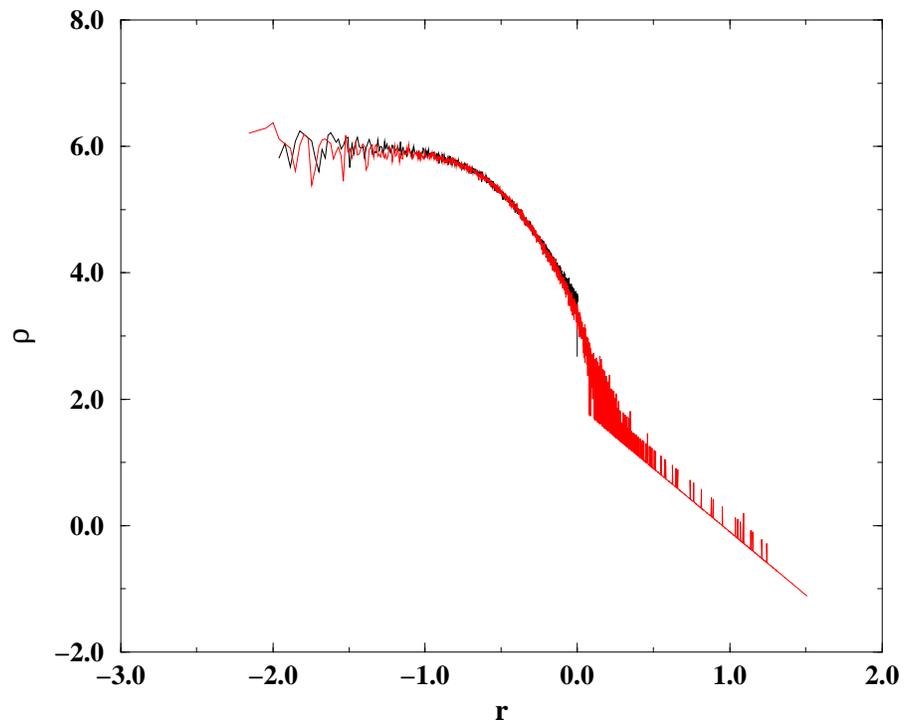,width=12.0cm}
       \vspace{0.5cm}
\label{fig6} \caption{Mass density of the system as a function of
the distance from the center (in logarithmic scales and in model
units), in the case of the system without the black hole, at t=0
(black line) and at t=10 t$_c$ (red line).}
\end{figure}

\newpage
\authorrunninghead{R. Capuzzo--Dolcetta et al.}
\titlerunninghead{On the use of MIMD--SIMD platform}
\begin{figure}[ht]
       \vspace{0.0cm}
       \noindent
       \psfig{file=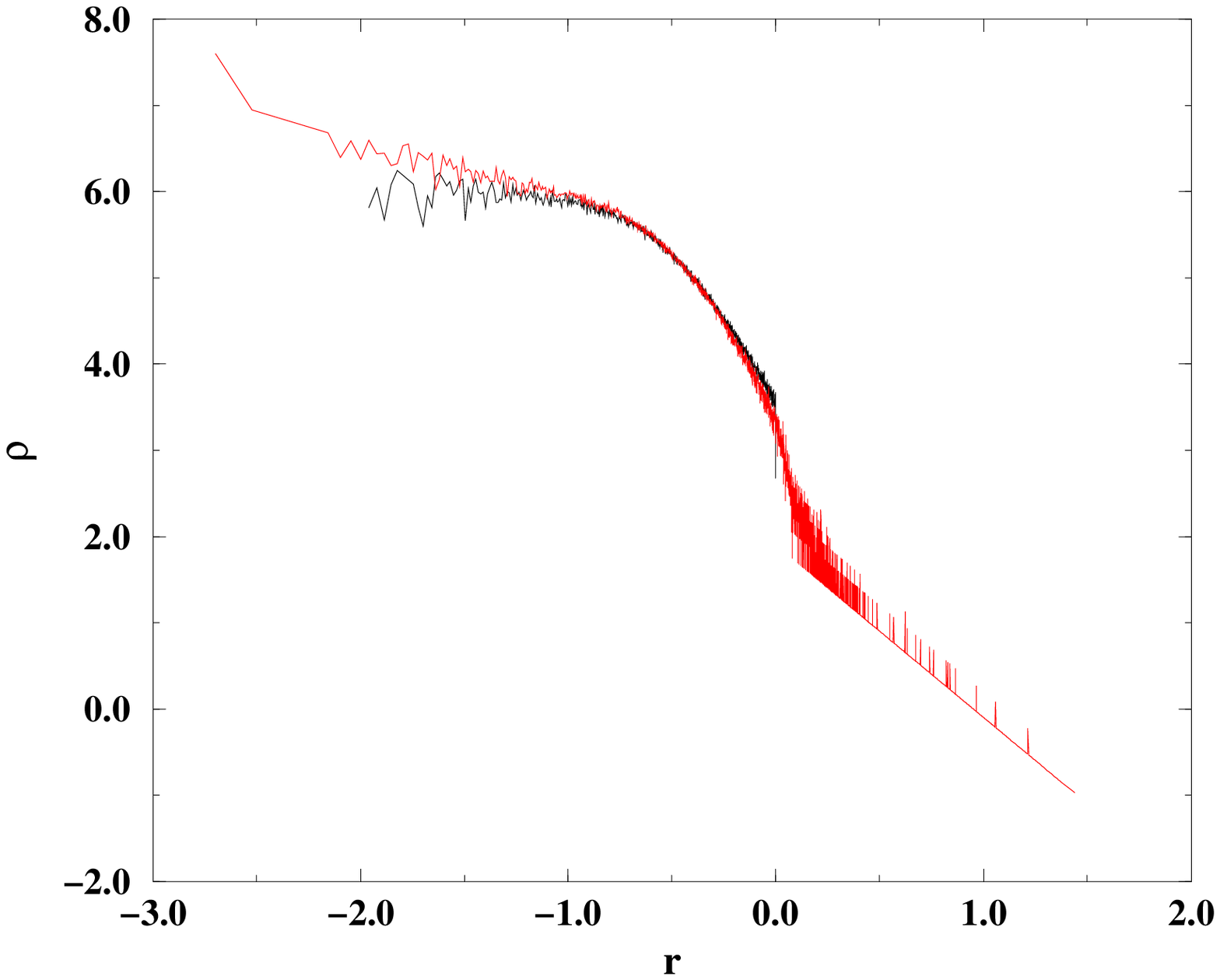,width=12.0cm}
       \vspace{0.5cm}
\label{fig7} \caption{Mass density of the system as a function of
the distance from the center (in logarithmic scales and in model
units) in the case $m_{BH}$=0.02 at t=0 (black line) and at t=10
t$_c$ (red line).}
\end{figure}

\newpage
\authorrunninghead{R. Capuzzo--Dolcetta et al.}
\titlerunninghead{On the use of MIMD--SIMD platform}
\begin{figure}[ht]
       \vspace{0.0cm}
       \noindent
       \psfig{file=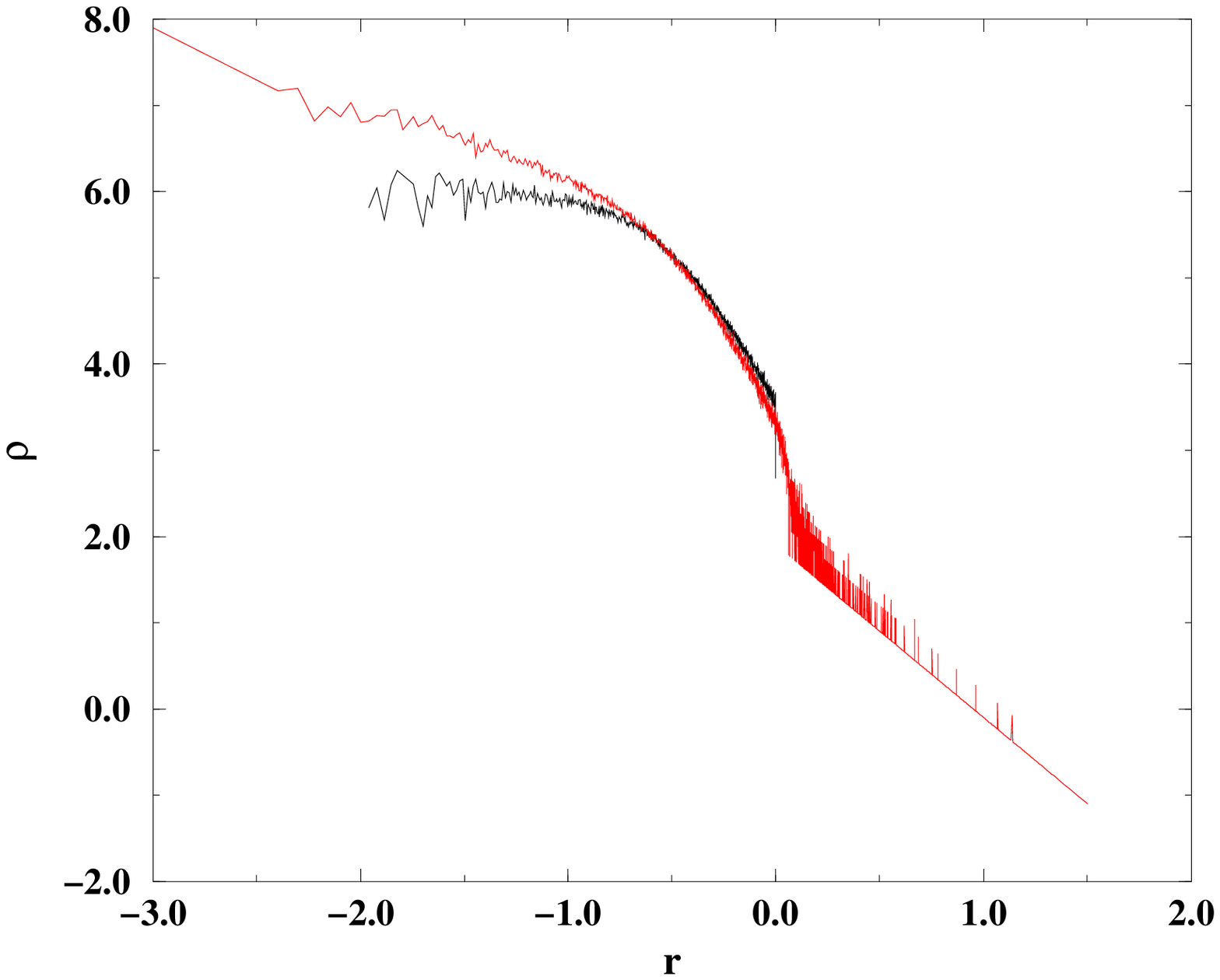,width=12.0cm}
       \vspace{0.5cm}
\label{fig8} \caption{Mass density of the system as a function of
the distance from the center (in logarithmic scales and in model
units) in the case $m_{BH}$=0.1 at t=0 (black line) and at t=10
t$_c$ (red line).}
\end{figure}

\newpage
\authorrunninghead{R. Capuzzo--Dolcetta et al.}
\titlerunninghead{On the use of MIMD--SIMD platform}
\begin{figure}[ht]
       \vspace{0.0cm}
       \noindent
       \psfig{file=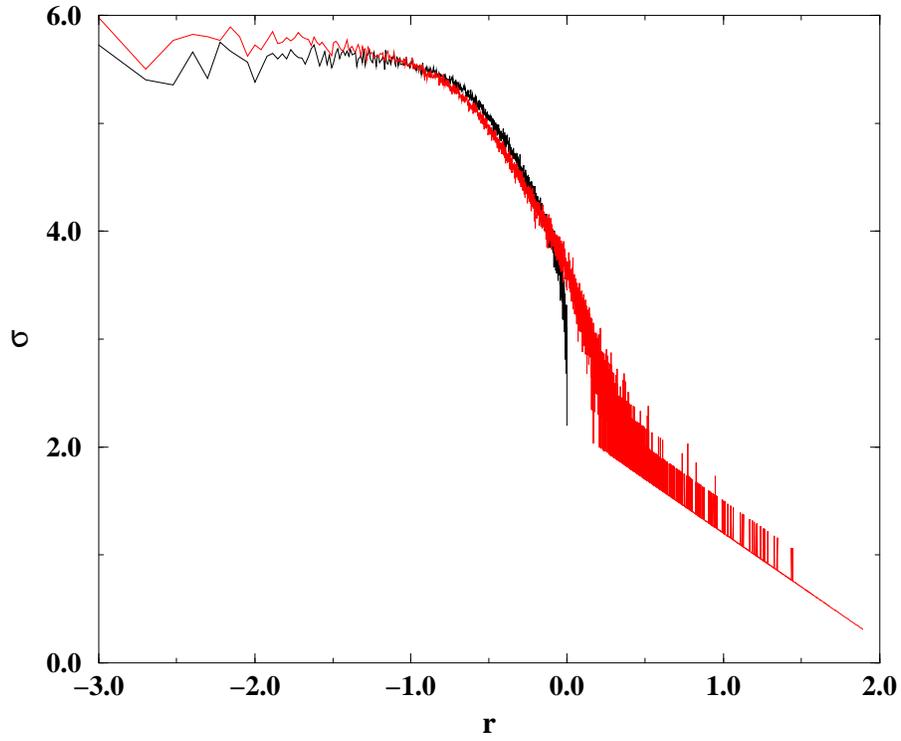,width=12.0cm}
       \vspace{0.5cm}
\label{fig9} \caption{Projected mass density of the system as a
function of the distance from the black--hole (in the case
$m_{BH}$=0.1) at t=0 (black line) and at t=10 t$_c$ (red line).}
\end{figure}

\newpage
\authorrunninghead{R. Capuzzo--Dolcetta et al.}
\titlerunninghead{On the use of MIMD--SIMD platform}
\begin{figure}[ht]
       \vspace{0.0cm}
       \noindent
       \psfig{file=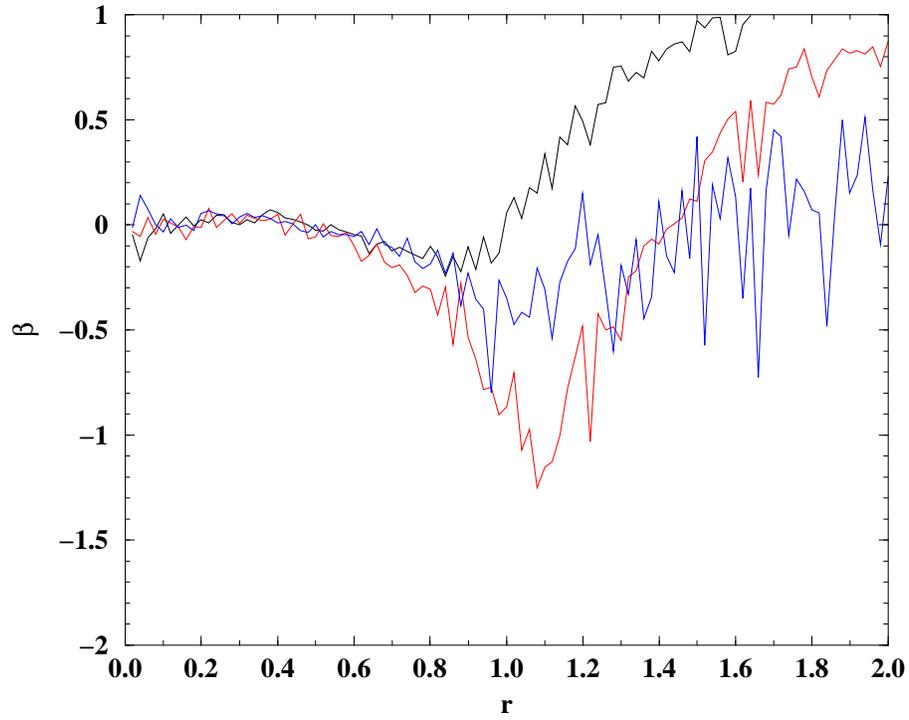,width=12.0cm}
       \vspace{0.5cm}
\label{fig10} \caption{ Anisotropy parameter $\beta$ (in model
units) for the $\theta$ component of the velocity for the system
without the black hole at t=0.3 t$_c$ (black line), t=1 t$_c$
(red line), t=10 t$_c$ (blue line).}
\end{figure}

\newpage
\authorrunninghead{R. Capuzzo--Dolcetta et al.}
\titlerunninghead{On the use of MIMD--SIMD platform}
\begin{figure}[ht]
       \vspace{0.0cm}
       \noindent
       \psfig{file=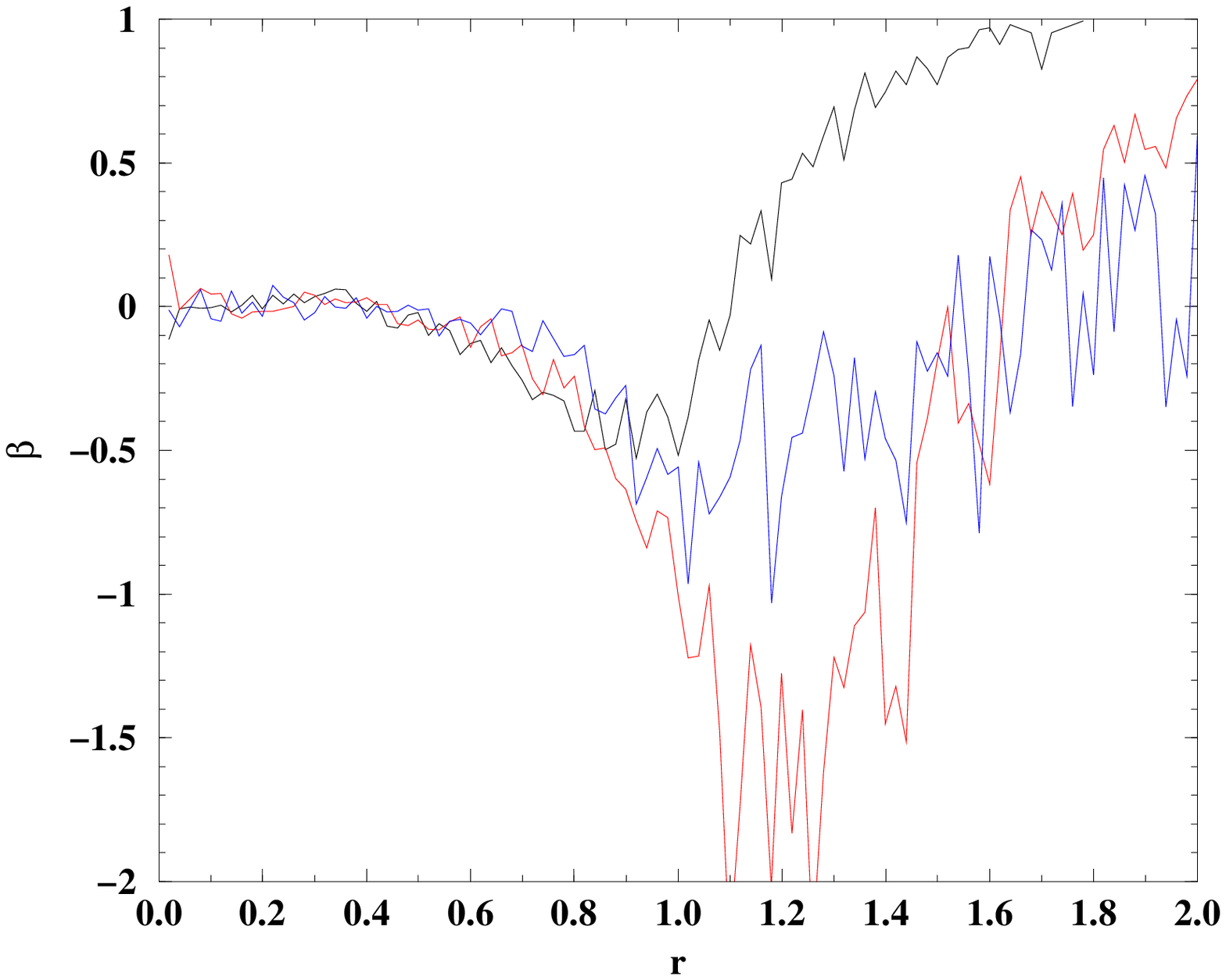,width=12.0cm}
       \vspace{0.5cm}
\label{fig11} \caption{Anisotropy parameter $\beta$ (in model
units) for the $\theta$ component of the velocity for the system
with $m_{BH}$=0.02 at t=0.2 t$_c$ (black line), t=1 t$_c$ (red
line), t=10 t$_c$ (blue line).}
\end{figure}

\newpage
\authorrunninghead{R. Capuzzo--Dolcetta et al.}
\titlerunninghead{On the use of MIMD--SIMD platform}
\begin{figure}[ht]
       \vspace{0.0cm}
       \noindent
       \psfig{file=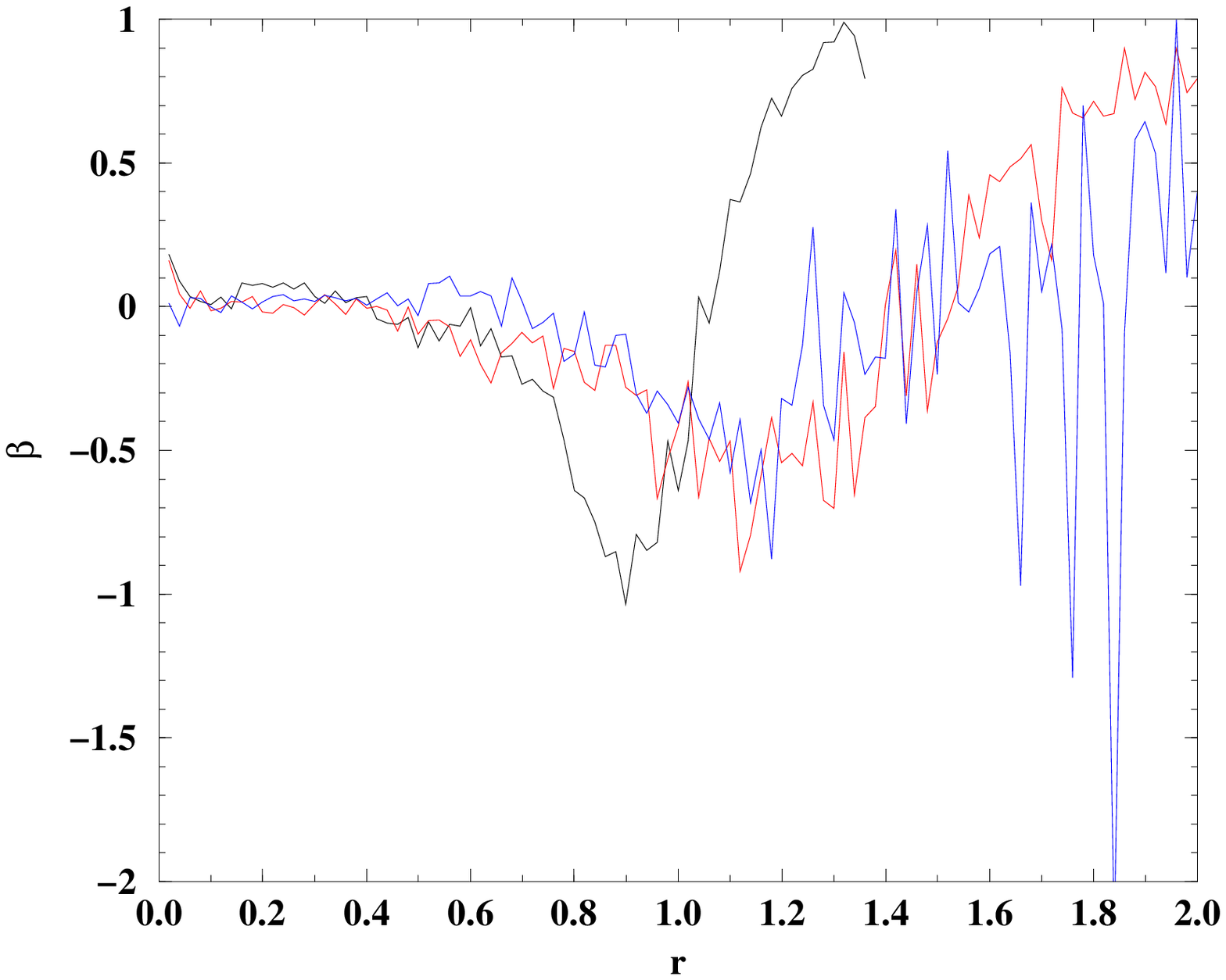,width=12.0cm}
       \vspace{0.5cm}
\label{fig12} \caption{Anisotropy parameter $\beta$ (in model
units) for the $\theta$ component of the velocity for the system
with $m_{BH}$=0.1 at t=0.2 t$_c$ (black line), t=1 t$_c$ (red
line), t=10 t$_c$ (blue line).}
\end{figure}


\end{article}
\end{document}